# Extending the UXR Point of View Pyramid: A Generative AI–Augmented Methodology for Human-Centred AI Systems

**Festus Fatai Adedoyin,**
School of Computing and Engineering, Bournemouth University Poole, UK, fadedoyin@bournemouth.ac.uk
**Huseyin Dogan**
School of Computing and Engineering, Bournemouth University Poole, UK, hdogan@bournemouth.ac.uk
**Melike Akca**
School of Computing and Engineering, Bournemouth University Poole, UK, makca@bournemouth.ac.uk
**Abiodun Adedeji**
School of Computing and Engineering, Bournemouth University Poole, UK, adedejia@bournemouth.ac.uk

**Abstract**

Rising household debt and cost-of-living pressures in the United Kingdom have intensified the role of AI-driven financial technologies in mediating credit assessment, repayment structuring, and debt support services. These systems increasingly shape consequential financial decisions, yet they operate within complex socio-technical environments characterised by regulatory constraint, algorithmic opacity, and heightened vulnerability risk. User Experience Research (UXR) Points of View (PoVs) are critical in translating heterogeneous research evidence into strategic direction for product and governance decisions. However, the existing UXR PoV framework was not designed for AI-mediated financial systems where interpretability, fairness, and accountability are central. This paper extends the UXR PoV pyramid into an AI-augmented methodological framework for Human-Centred AI debt management technologies in the UK financial services context. We formalise (1) an AI-Augmented PoV Pyramid, (2) a structured prompt architecture for synthesis and hypothesis generation, and (3) an AI-enabled Playbook Card system that embeds Generative AI into UXR workflows while preserving traceability and ethical oversight. Generative AI is positioned not as an analytic authority, but as an epistemic support mechanism subject to human validation and regulatory awareness. By grounding the framework in debt management technologies, including affordability assessment, repayment planning, and financial stress prediction systems, this work advances UXR methodology for high-stakes financial AI environments and contributes to the evolution of responsible, AI-powered UXR practice within the CHI community.

## 1 INTRODUCTION

User Experience Research (UXR) faces a persistent translational challenge: converting complex empirical evidence into insights that credibly inform design, governance, and strategy. The UXR Point of View (PoV) framework addresses this by structuring research into a coherent narrative architecture for multidisciplinary decision-making [1]. However, the epistemic conditions of PoV construction have shifted, particularly in high-stakes financial contexts.

In the United Kingdom, rising household debt and cost-of-living pressures have increased reliance on AI-enabled financial systems. Algorithmic models now mediate credit scoring, affordability assessment, repayment structuring, financial stress prediction, and automated debt support. These probabilistic, often opaque systems directly affect financially vulnerable consumers. Human-Centred AI (HCAI) therefore requires reliability, transparency, and governance mechanisms that preserve meaningful human oversight [2].

Generative AI (GenAI) adds capabilities for synthesis, reframing, summarisation, and scenario modelling. While it can enhance research workflows, unstructured use risks epistemic outsourcing, reduced traceability, and amplification of structural bias [3,4]. In debt management technologies, such risks are material: persuasive but weakly grounded narratives may shape repayment policies, vulnerability classification, and user treatment strategies.

This paper addresses the methodological tension: how can GenAI be systematically integrated into the UXR PoV pyramid to support Human-Centred AI debt management technologies in the UK while preserving accountability and rigour? Grounded in UK financial services and AI-mediated debt support systems, it extends the PoV framework into an AI-Augmented UXR PoV Pyramid, formalising structured mechanisms for prompt architecture, hypothesis generation, and operational playbooks.

## 2. RELATED WORK

Human-Centred AI conceptualises AI systems as socio-technical infrastructures embedded in institutional, regulatory, and lived contexts [2]. In UK financial services, AI increasingly mediates affordability assessment, repayment restructuring, risk segmentation, and vulnerability detection, shaping repayment flexibility, credit rehabilitation access, and escalation procedures.

Algorithmic systems can reproduce or amplify structural inequalities when trained on historically biased data. Governance frameworks are therefore essential for legitimacy and accountability in AI-enabled finance [5]. UXR PoVs must extend beyond articulating user needs to incorporate regulatory awareness, ethical scrutiny, and analysis of algorithmic consequences. In debt management, the PoV aligns user welfare with institutional compliance and risk governance.

Explainable AI research positions interpretability as central to trust and institutional legitimacy. In debt systems subject to regulatory oversight, explainability is critical. UXR PoVs function as epistemic mediators, synthesising qualitative accounts of financial stress, behavioural signals, quantitative risk indicators, and organisational constraints into structured strategic narratives [6].

GenAI expands this mediation capacity but, without safeguards, may obscure evidence provenance or reinforce dominant financial narratives. In UK debt management contexts, structured integration is therefore essential to maintain evidentiary integrity and methodological rigour.

## 3. METHOD

This study employed a Generative AI-augmented UXR PoV workshop methodology to extend the UXR PoV Playbook into UK AI-driven debt management systems [1,7]. It integrates practitioner expertise, structured prompt architecture, and regulatory governance principles to produce design-actionable and compliance-defensible outputs for affordability assessment and repayment restructuring technologies [2,8].

Rather than adopting a single analytical lens, the study applied a multi-perspective synthesis combining Human-Centred AI principles [2], financial vulnerability frameworks [8], regulatory defensibility requirements, and situational decision-cycle models (Endsley, 1995; Klein, 1998) [5,9,10]. This pluralistic design reflects the complexity of AI deployment in regulated financial services, where behavioural insight, institutional accountability, data governance, and harm mitigation intersect [9,6].

The methodology followed four structured stages aligned with the AI-powered UXR PoV framework (Dogan et al., 2025): GenAI-supported practice mapping and hypothesis generation; foundational governance and stakeholder road mapping; insight generation and debt-focused Play development; and PoV narrative construction and stakeholder translation.

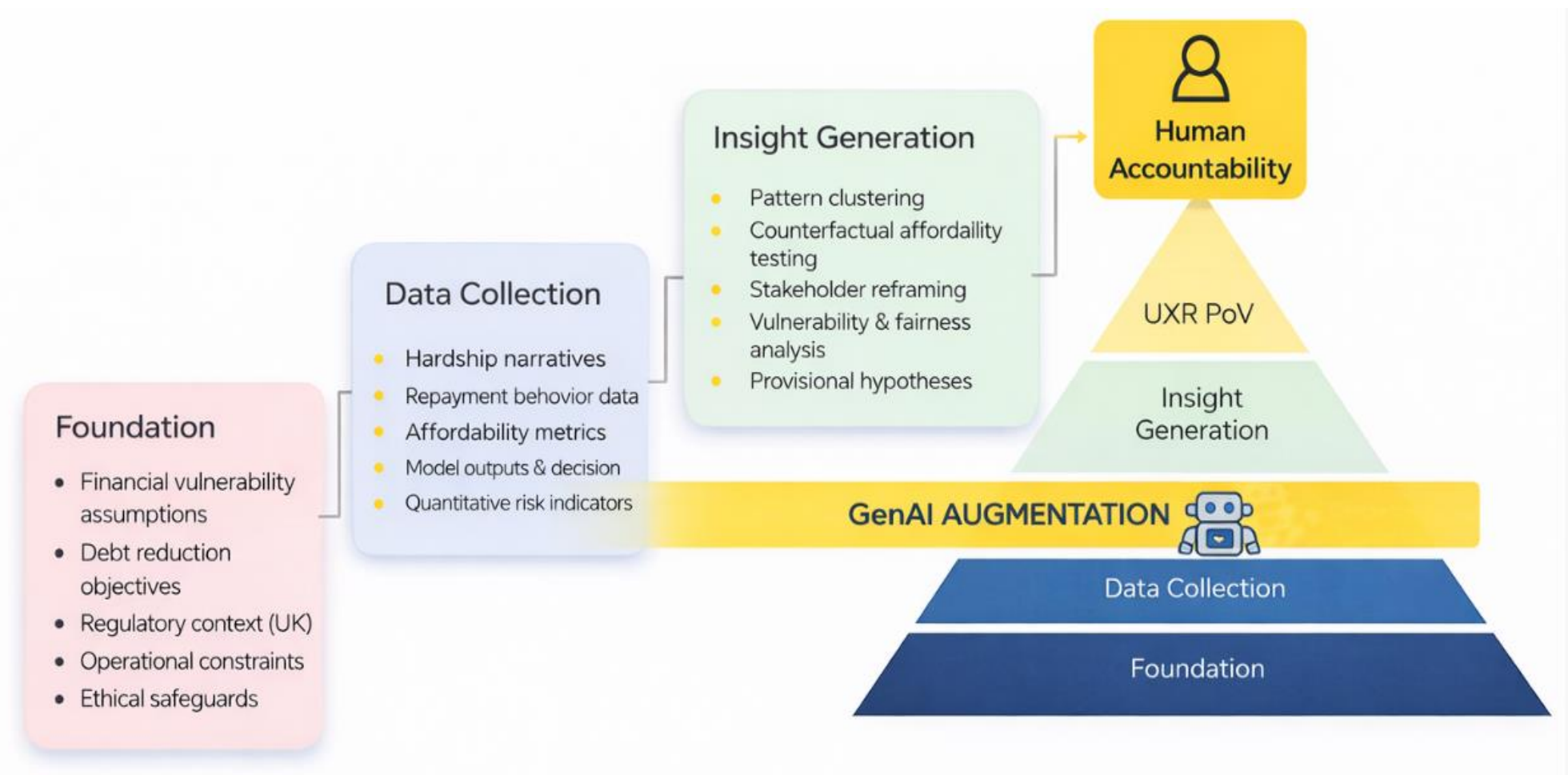

*Figure 1. Four-Stage AI-Augmented UXR PoV Research Process for UK Debt Management.*

**3.1 Evidence Sources and Data Preparation**

The primary corpus comprised structured workshop artefacts and domain-specific operational knowledge from UK debt management contexts. These included documented AI-assisted research practices (e.g., summarisation of hardship interviews, clustering of repayment behaviour), affordability modelling reflections, compliance documentation workflows, and governance review procedures.

Participants contributed experiential knowledge derived from regulated financial environments, particularly around affordability assessment systems and repayment restructuring tools. These inputs were thematically organised prior to AI interaction under the following categories:

- Financial vulnerability and hardship narratives
- Repayment behaviour patterns
- Affordability modelling constraints
- Regulatory and compliance requirements
- Transparency and explainability challenges

This structured preparation ensured that AI-supported synthesis operated on curated thematic inputs rather than unstructured workshop transcripts. All materials were provided to the GenAI system alongside the UXR PoV framework and Playbook documentation to enable cross-framework integration.

**3.2 Procedures and Prompts**

**Stage 1: GenAI-Supported Practice Mapping and Hypothesis Generation**

The first stage aimed to surface recurring analytical patterns in AI use within UK debt research and translate them into defensible hypotheses regarding risk, governance, and design intervention.

GenAI was directed to synthesise existing practices, identify epistemic gaps, and propose structured hypotheses linking AI-supported insight generation to regulatory defensibility and vulnerability protection outcomes.

Prompts submitted included:

- "Analyse current uses of Generative AI in UK debt management research and identify recurring analytical patterns."
- "Identify risks of unstructured AI use in affordability assessment and hardship analysis."
- "Cluster these insights within the UXR PoV layers (Foundation → Data Collection → Insight Generation → PoV)."
- "Generate testable hypotheses linking AI-supported workflows to transparency, bias mitigation, and regulatory defensibility outcomes."

AI-generated hypotheses were treated as provisional and required human validation against domain knowledge and governance requirements.

**Stage 2: Foundational Planning and Stakeholder Road Mapping**

Stage 2 established contextual and governance grounding. GenAI was instructed to simulate stakeholder mapping, vulnerability impact analysis, and regulatory alignment planning. The objective was to articulate a shared roadmap linking research goals, institutional accountability, and harm mitigation safeguards.

Prompts submitted included:

- "Describe how financially vulnerable individuals experience AI-assisted affordability assessments."
- "Identify cognitive, emotional, and financial risks introduced by opaque AI decision systems."
- "Map primary and secondary stakeholders in UK AI-driven debt management, including regulators."
- "Define evidential standards, transparency requirements, and accountability mechanisms for each stakeholder group."
- "Create a step-by-step project plan linking research objectives, compliance safeguards, and stakeholder engagement using mixed methods (qualitative, quantitative, AI-supported synthesis)."

Outputs from this stage defined governance checkpoints and clarified validation pathways for subsequent insight development.

**Stage 3: Insight Generation and Debt-Focused Play Development**

Stage 3 operationalised the Building Blocks model, Foundation, Data Collection, Insight Generation, and PoV translation to produce structured debt-focused Plays.

GenAI was used iteratively to cluster empirical and experiential insights, generate structured hypotheses, and translate validated findings into design- and governance-oriented artefacts.

Prompts submitted included:

- “Cluster hardship and repayment data into themes of transparency, fairness, motivation, and trust.”
- “Compare repayment outcomes across vulnerability segments and identify potential bias amplification.”
- “Generate counterfactual scenarios stress-testing affordability thresholds.”
- “Produce Play Card content linking regulatory requirement, empirical pattern, and design or governance intervention.”
- “Use example Play Cards as a structural reference when generating outputs.”

Each Play articulated:

- The identified financial or behavioural challenge
- Supporting evidence
- Governance or design intervention
- Expected impact on bias mitigation, transparency, or vulnerability protection

A structured hypothesis validation workflow required triangulation across qualitative hardship narratives, quantitative repayment metrics, and compliance standards before acceptance.

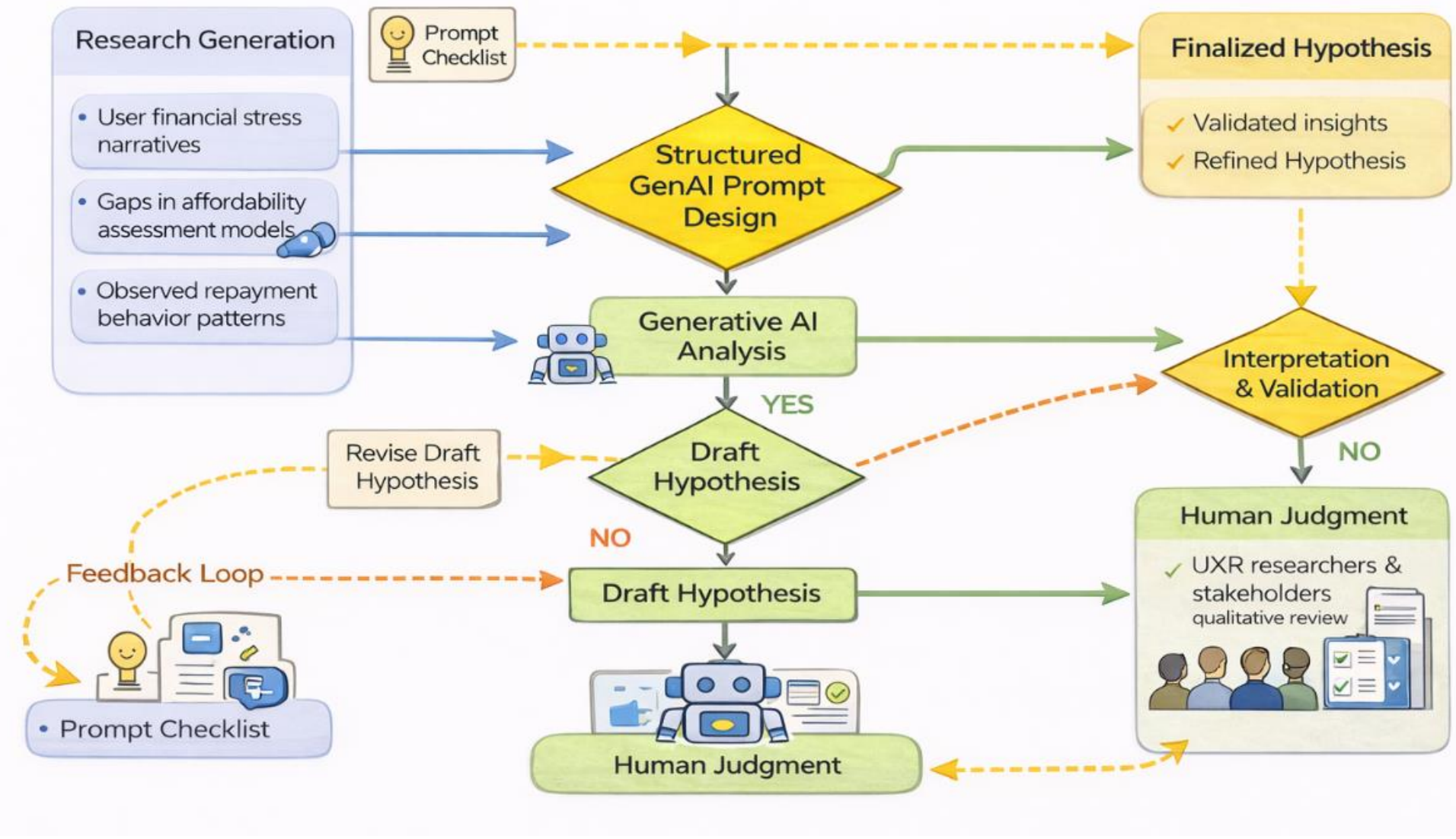

*Figure 2. Structured Hypothesis Generation and Validation Workflow.*

**Stage 4: PoV Narrative Construction and Stakeholder Communication**

The final stage translated validated Plays into coherent, stakeholder-specific UXR PoV narratives suitable for financial institutions, compliance officers, product teams, and governance leaders. GenAI supported iterative phrasing and structural refinement; interpretive authority remained with the research team.

Prompts submitted included:

- “Generate a debt-management-specific PoV narrative aligned with UK regulatory defensibility.”
- “Draft compliance-focused PoV narratives for governance stakeholders.”
- “Refine the PoV statement for transparency, fairness, and vulnerability-sensitive language.”
- “Summarise Play Card insights into concise executive-level stakeholder messages.”
- “Ensure consistency with financial vulnerability protection principles.”

All outputs underwent expert review to ensure conceptual coherence, regulatory alignment, and evidential traceability.

Together, these four stages constitute a structured AI-augmented UXR PoV methodology tailored to UK debt management systems. The approach formalises Generative AI as a synthesis accelerator embedded within accountable, human-governed research workflows.

## 4 RESULTS

This study produced a structured set of findings demonstrating how Generative AI (GenAI) can be systematically embedded within the UXR Point of View (PoV) pyramid to support Human-Centred AI debt management technologies in the UK financial services context. Results are organised according to the four-stage AI-augmented UXR PoV framework:

(1) GenAI-supported practice mapping and hypothesis generation,
(2) foundational governance and stakeholder road mapping,
(3) insight generation and AI-augmented Playbook Card development, and
(4) stakeholder-aligned PoV narrative construction.

Across these stages, Generative AI was positioned as an epistemic support mechanism rather than an analytic authority. Outputs were iteratively validated against domain expertise, regulatory standards (FCA Consumer Duty and vulnerability guidance), and financial governance requirements. The findings illustrate how structured GenAI integration can enhance synthesis, traceability, and defensibility within high-stakes financial AI environments.

### 4.1 Stage 1: GenAI-Supported Practice Mapping and Hypothesis Generation

In the first stage, Generative AI was used to analyse structured workshop artefacts and operational knowledge from UK AI-enabled debt management contexts. These included affordability assessment practices, hardship analysis workflows, repayment restructuring processes, and governance review mechanisms.

Through structured prompt architecture (exploratory, contrastive, counterfactual, and ethical interrogation prompts), recurring analytical patterns were identified and translated into defensible hypotheses regarding transparency, bias mitigation, and regulatory defensibility. Ten recurring themes emerged: narrative compression risks in hardship summarisation; hidden assumptions in affordability thresholds; income volatility bias in gig-economy and irregular earners; automation-induced overconfidence in repayment predictions; regulatory traceability gaps in AI-supported synthesis; stakeholder-driven reframing of vulnerability narratives; Interpretability tensions between risk teams and UX teams; Overreliance on historical repayment data; Lack of structured ethical stress testing; Narrative drift during executive translation. From these themes, ten structured hypotheses were generated:

**H1:** Structured prompt architecture reduces narrative compression bias in hardship analysis.
**H2:** Counterfactual prompting improves identification of affordability threshold inequities.
**H3:** Income smoothing simulations mitigate volatility bias in gig-economy assessments.
**H4:** Explicit uncertainty framing reduces automation-induced overconfidence in repayment predictions.
**H5:** Evidence-tagging prompts improve audit traceability of AI-supported synthesis.
**H6:** Stakeholder reframing checkpoints reduce vulnerability narrative drift.
**H7:** Interpretability prompts enhance cross-functional alignment between UX and risk teams.
**H8:** Bias interrogation prompts reduce historical data amplification in risk modelling.
**H9:** Ethical stress testing improves identification of hardship detection blind spots.
**H10:** Structured hypothesis validation workflows enhance regulatory defensibility of PoVs.

Each hypothesis was treated as provisional and subjected to human validation against domain practice and compliance requirements before inclusion in the framework.

### 4.2 Stage 2: Foundational Governance and Stakeholder Road Mapping

The second stage established governance grounding and stakeholder alignment. GenAI supported the articulation of cognitive, emotional, financial, and regulatory risks introduced by AI-mediated debt systems.

Five primary stakeholder groups were mapped:

| Stakeholder | Role | Core Need | Primary Challenge |
|---|---|---|---|
| Financially Vulnerable Customers | End-users | Fair, comprehensible affordability assessments | Opaque algorithmic decisions affecting repayment burden |
| Compliance & Risk Officers | Governance gatekeepers | Regulatory defensibility and audit traceability | Balancing automation efficiency with Consumer Duty obligations |
| UX Researchers | Translators | Evidence-grounded PoVs for financial products | Preventing narrative drift in AI-supported synthesis |
| Data Scientists / AI Engineers | Model builders | Robust, interpretable modelling logic | Mitigating bias without degrading predictive performance |
| Product & Strategy Leads | Decision-makers | Scalable and competitive financial solutions | Aligning user welfare with institutional risk appetite |

This stakeholder mapping surfaced systemic tensions:

- Transparency vs. competitive model secrecy
- Automation efficiency vs. vulnerability protection
- Risk minimisation vs. consumer fairness
- Narrative persuasion vs. evidential traceability

These tensions informed the development of governance checkpoints embedded within subsequent Playbook Cards and PoV narratives.

### 4.3 Stage 3: AI-Augmented Insight Generation and Playbook Card Development

In the third stage, validated hypotheses were operationalised into AI-Augmented Playbook Cards tailored to UK debt management technologies. Each card includes Purpose, Structured AI prompt template, Human validation step, Ethical checkpoint, Expected governance impact. Ten core Playbook Cards were developed:

| Card # | Title | Issue Type | Governance Intervention |
|---|---|---|---|
| 1 | Hardship Narrative Compression | Summary bias | Evidence-tagged AI prompts |
| 2 | Affordability Threshold Stress Testing | Structural inequity | Counterfactual income simulations |
| 3 | Income Volatility Bias | Data skew | Alternative smoothing models |
| 4 | Automation Overconfidence | Model opacity | Explicit uncertainty framing |
| 5 | Untraceable AI Synthesis | Audit risk | Source citation enforcement |
| 6 | Stakeholder Narrative Drift | Persuasive reframing | Traceability validation checkpoint |
| 7 | UX–Risk Interpretability Gap | Cross-functional misalignment | Interpretability translation prompts |
| 8 | Historical Bias Amplification | Structural inequity | Bias interrogation prompts |
| 9 | Ethical Stress Testing | Harm blind spots | Counterfactual vulnerability scenarios |
| 10 | Regulatory Defensibility Mapping | Governance misalignment | Hypothesis validation workflow |

These cards function as boundary objects between UX, compliance, and AI governance teams. They formalise defensible integration of GenAI within regulated financial research workflows while preserving accountability.

### 4.4 Stage 4: Stakeholder-Aligned PoV Narrative Construction

In the final stage, validated Plays were synthesised into stakeholder specific UXR PoV narratives. GenAI supported iterative phrasing, structural coherence, and scenario articulation, while interpretive authority remained human-led. Four cross cutting PoV pillars emerged:

1. **Traceability**: Every AI-supported insight must be auditable and source-linked.
2. **Fairness Sensitivity**: Affordability modelling must account for income volatility and structural disadvantage.
3. **Interpretability**: Model outputs must be communicable across risk, UX, and compliance teams.
4. **Ethical Stress Testing**: Counterfactual simulation must be embedded before deployment.

Stakeholder-aligned PoV Template emphases included:

| Stakeholder | Core Need | Primary Challenge | PoV Narrative Focus | GenAI-Supported Contribution |
|---|---|---|---|---|
| Financially Vulnerable Customers | Transparent, fair affordability and repayment decisions | Opaque AI-driven assessments that may increase repayment burden | Clear, comprehensible affordability explanations; safeguards against structural disadvantage; reduced undue hardship | Scenario simulation of alternative affordability rules; clarity testing of decision explanations; counterfactual hardship prompts |
| Compliance & Risk Leaders | Regulatory defensibility and audit traceability | Balancing automation efficiency with FCA Consumer Duty obligations | Evidential traceability across research, modelling, and decision logic; defensible AI-supported workflows | Evidence-tagging prompts; structured hypothesis validation workflow; compliance reframing checks |
| UX Researchers | Evidence-grounded PoVs that influence product direction | Narrative drift and overreliance on persuasive AI summarisation | Structured prompt architecture; defensible synthesis; traceable translation from hardship data to PoV | Exploratory and contrastive prompt templates; narrative drift detection prompts; structured clustering mechanisms |
| AI Engineers / Data Scientists | Robust and interpretable predictive models | Mitigating bias while preserving predictive performance | Interpretable modelling logic aligned with fairness and vulnerability sensitivity | Bias interrogation prompts; counterfactual income simulations; uncertainty-framing outputs |
| Product & Strategy Leaders | Competitive, scalable financial innovation | Aligning commercial risk appetite with vulnerability safeguards | Strategic PoV aligning innovation, governance, and consumer protection | Executive reframing prompts; stakeholder translation templates; ethical stress-testing summaries |

## 5 DISCSUSSION

This research extends the UXR Point of View (PoV) framework into regulated financial AI environments by embedding Generative AI within structured methodological safeguards. A principal contribution is the operationalisation of Human-Centred AI (HCAI) in high-stakes financial systems. HCAI prioritises reliability, transparency, and meaningful human oversight over automation-centric optimisation [2,12]. Calibrated trust and appropriate human control are central to responsible human-AI interaction [11]. By positioning Generative AI as an augmentation layer across the PoV pyramid rather than as an epistemic authority, the framework aligns with guidance advocating structured oversight in AI-supported decision environments [5]. In UK debt management contexts, where algorithmic affordability assessments directly affect vulnerable consumers, governance-aware prompt architectures and validation checkpoints strengthen transparency and defensibility in line with responsible AI deployment [8].

A second contribution addresses algorithmic bias and fairness risks in financial systems. AI models may reproduce structural inequalities embedded in historical credit and repayment data [9,13]. Interpretability research further emphasises that transparency must encompass communicative clarity, auditability, and explanation quality not solely model internals [6,14]. By embedding counterfactual prompting, bias interrogation, and evidence-tagging within PoV construction, the framework moves fairness evaluation upstream into research synthesis rather than confining it to post hoc auditing. This approach aligns with governance scholarship advocating lifecycle oversight, documentation, and socio-technical evaluation of AI systems [15,16]. Accordingly, UXR artefacts are reframed as governance infrastructure mediating between financial modelling, regulatory compliance, and consumer protection.

Several limitations remain. Generative AI systems are probabilistic and vulnerable to hallucination, bias amplification, and persuasive narrative drift [17,18]. Structured prompting and validation workflows mitigate but do not eliminate epistemic uncertainty or latent representational bias. Effectiveness also depends on organisational maturity, cross-functional alignment, and access to high-quality hardship and repayment data conditions unevenly distributed across financial institutions [19]. Moreover, while this study formalises defensible AI-supported UXR workflows, it does not yet provide longitudinal validation of downstream outcomes, such as measurable improvements in fairness metrics, repayment stability, or regulatory audit performance. Future research should therefore conduct applied case evaluations to assess impacts on decision quality, consumer protection, and institutional accountability within operational debt management systems.

## 6 CONCLUSION AND FUTURE WORK

This paper extends the UXR PoV Playbook by embedding Generative AI within structured methodological safeguards tailored to UK debt management technologies. The AI-Augmented PoV framework integrates the pyramid model, prompt architecture, hypothesis validation workflows, and defensive plays to ensure that AI augments rather than replaces human interpretive judgment in regulated financial contexts.

AI-driven debt management systems in the UK operate under strict regulatory requirements, including FCA guidance and Consumer Duty obligations. UX researchers in these environments must construct Points of View that are strategically aligned, explainable, and ethically defensible. The AI-Augmented PoV Playbook provides structured mechanisms for integrating Generative AI into research workflows while preserving accountability. Adaptable building blocks support evidence-grounded synthesis, bias-aware hypothesis testing, and defensible stakeholder communication. In debt management contexts, this strengthens traceability between hardship indicators, repayment modelling, and design recommendations, while reducing narrative drift and ungrounded AI interpretations. Future work will focus on aligning AI-aware UXR planning with affordability modelling, vulnerability detection, and fairness evaluation in UK financial services.

As a limitation, generative AI systems remain probabilistic and vulnerable to bias and hallucination. Structured prompting reduces but does not eliminate these risks. In debt management systems affecting financially vulnerable users, human oversight remains essential. Implementation will vary depending on organisational maturity, governance structures, and data quality. The framework provides methodological discipline but cannot control broader socio-technical constraints. Validation is ongoing through case studies in AI-enabled financial service teams, including debt and affordability contexts. Evaluations assess improvements in traceability, interpretive rigour, and regulatory defensibility. Practitioner workshops and expert review from AI governance specialists will further refine the framework for responsible AI-assisted UXR in regulated financial systems.